\documentclass[pra, onecolumn,notitlepage, longbibliography, floatfix, superscriptaddress]{revtex4-1}
\usepackage{dcolumn}    
\usepackage{bm} 
\usepackage{graphicx}
\usepackage{amsmath}    
\usepackage{enumitem}
\usepackage{latexsym}
\usepackage{amsfonts}   
\usepackage{amssymb}
\usepackage{array}      
\usepackage{epsfig}
\usepackage[dvipsnames]{xcolor}
\usepackage{color}
\usepackage[colorlinks=true,linkcolor=blue,urlcolor=blue,citecolor=blue,pdfusetitle]{hyperref}
\usepackage{hyperref}
\usepackage[capitalize]{cleveref}
\usepackage{physics}
\usepackage{bbold}

\newcommand{\s}{\sigma}
\newcommand{\cf} {cf.~}
\newcommand{\eq}{Eq.~}
\newcommand{\eqs}{Eqs.~}

\newcommand{\fig}{Fig.~}

\begin{document}

\title{Microscopic biasing of discrete-time quantum trajectories}

\author{Dario Cilluffo}
\email{Corresponding author:dario.cilluffo@unipa.it}
\affiliation{Universit$\grave{a}$  degli Studi di Palermo, Dipartimento di Fisica e Chimica - Emilio Segr\`e, via Archirafi 36, I-90123 Palermo, Italy}
\affiliation{NEST, Istituto Nanoscienze-CNR, Piazza S. Silvestro 12, 56127 Pisa, Italy}
\affiliation{Institut f\"ur Theoretische Physik, Universit\"at T\"ubingen, Auf der Morgenstelle 14, 72076 T\"ubingen, Germany}
\author{Giuseppe Buonaiuto}
\affiliation{Institut f\"ur Theoretische Physik, Universit\"at T\"ubingen, Auf der Morgenstelle 14, 72076 T\"ubingen, Germany}
\author{Igor Lesanovsky}
\affiliation{School of Physics and Astronomy, University of Nottingham, Nottingham, NG7 2RD, United Kingdom}
\affiliation{Centre for the Mathematics and Theoretical Physics of Quantum Non-equilibrium Systems, University of Nottingham, Nottingham, NG7 2RD, United Kingdom}
\affiliation{Institut f\"ur Theoretische Physik, Universit\"at T\"ubingen, Auf der Morgenstelle 14, 72076 T\"ubingen, Germany}
\author{Angelo Carollo}
\affiliation{Universit$\grave{a}$  degli Studi di Palermo, Dipartimento di Fisica e Chimica - Emilio Segr\`e, via Archirafi 36, I-90123 Palermo, Italy}
\affiliation{Radiophysics Department, National Research Lobachevsky State University of Nizhni Novgorod, 23 Gagarin Avenue, Nizhni Novgorod 603950, Russia}
\author{Salvatore Lorenzo}
\affiliation{Universit$\grave{a}$  degli Studi di Palermo, Dipartimento di Fisica e Chimica - Emilio Segr\`e, via Archirafi 36, I-90123 Palermo, Italy}
\author{G. Massimo Palma}
\affiliation{Universit$\grave{a}$  degli Studi di Palermo, Dipartimento di Fisica e Chimica - Emilio Segr\`e, via Archirafi 36, I-90123 Palermo, Italy}
\affiliation{NEST, Istituto Nanoscienze-CNR, Piazza S. Silvestro 12, 56127 Pisa, Italy}
\author{Francesco Ciccarello}
\affiliation{Universit$\grave{a}$  degli Studi di Palermo, Dipartimento di Fisica e Chimica - Emilio Segr\`e, via Archirafi 36, I-90123 Palermo, Italy}
\affiliation{NEST, Istituto Nanoscienze-CNR, Piazza S. Silvestro 12, 56127 Pisa, Italy}
\author{Federico Carollo}
\affiliation{Institut f\"ur Theoretische Physik, Universit\"at T\"ubingen, Auf der Morgenstelle 14, 72076 T\"ubingen, Germany}


\begin{abstract}
We develop a microscopic theory for biasing the quantum trajectories of an open quantum system,
which renders rare trajectories typical.
To this end we consider a discrete-time quantum dynamics, where the open system collides sequentially with qubit probes which are then measured. A theoretical framework is built in terms of thermodynamic functionals in order to characterize its quantum trajectories (each embodied by a sequence of measurement outcomes). We show that the desired biasing is achieved by suitably modifying the Kraus operators describing the discrete open system dynamics. From a microscopical viewpoint and for short collision times, this corresponds to adding extra collisions which enforce the system to follow a desired rare trajectory. The above extends the theory of biased quantum trajectories from Lindblad-like dynamics to sequences of arbitrary dynamical maps, providing at once a transparent physical interpretation.
\end{abstract}
\date{\today}
\maketitle

\section{Introduction}
Controlling quantum systems typically requires coping with dissipative (open) nonequilibrium dynamics. As dissipation is due to the ineliminable effect of an environment, the system-environment coupling needs to be controlled, or even explicitly harnessed, as in dissipative quantum computing/state engineering \cite{KempePRA01, DFLidar, DFPalma} or preparation of decoherence-free steady states \cite{PhysRevLett.89.277901,BeigePRL00,GonzalezTudelaPRL15}.

At a fundamental level, open quantum dynamics result from single stochastic realizations called quantum trajectories \cite{Gardiner2004,breuerpetruccione}. In each of these, the open system, which is effectively continuously monitored by the “environment”, undergoes an overall non-unitary time-evolution interrupted, at random times, by quantum jumps. Each jump, which for atom-photon systems is in one-to-one correspondence with the irreversible emission and detection of a photon, causes a sudden change of the system state \cite{ZollerPRA1987,BrunPRA2000}.

\begin{figure}
	\includegraphics[scale=1.3,angle=0]{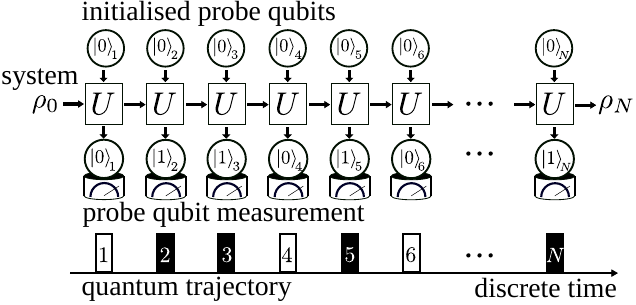}
	\caption{{\bf Quantum collision model.} The environment consists of a large collection of quantum probes, each modeled as a qubit with (orthonormal) computational basis $\{\ket{0},\ket{1}\}$. This environment is initialized in the state $\otimes_n\! \ket{0_n}$. The system, whose initial state is described by the density matrix $\rho_0$, collides with the probes one at a time, the $n$th collision being described by a pairwise unitary $U$ on the system and probe $n$. As the collision is complete (and before the system collides with probe $n+1$), probe $n$ is measured in the basis $\{\ket{k}\}$ with $k=0,1$. When the outcome $\ket{1}$ is detected, a quantum jump occurs (black square). The sequence of measurement outcomes uniquely defines a quantum trajectory. Summing over all possible realizations of the measurement provides, instead, the dynamics of the average system state $\rho_n$.}
	\label{fig1}
\end{figure}

In analogy with equilibrium thermodynamic ensembles \cite{touchette2009large,greiner2012thermodynamics}, the collection of quantum trajectories and their probability can be treated as a macroscopic (non-equilibrium) state. Each single realization can be thought of as a microstate, characterized by the number of occurred jumps, typically extensive with the observation time. The average properties of the system are determined by typical trajectories of the macroscopic state, while rare ones govern deviations from such behaviour \cite{10.1143/PTPS.184.304,PhysRevLett.111.120601,garaspects,Jack:2020aa}. 

Given the above scenario, {controlling the statistics of trajectories} is crucial: a major benefit would, e.g., be the possibility of engineering devices with the desired emission properties. This is yet a challenging task since changing the jumps statistics in fact entails turning rare trajectories into typical \cite{garrahan2010thermodynamics,carollo2018making,PhysRevLett.122.130605}. First progress along this line was recently made by showing that a preselected set of rare trajectories of a Markovian open quantum system described by a Lindblad master equation can always be seen as the typical realizations of an alternative (still Markovian) system \cite{carollo2018making}. Yet, the physical connection between the two systems (which may be radically different) is not straightforwardly interpreted.

This work approaches the problem of tailoring trajectories statistics from a much wider viewpoint in two main respects. On the one hand, we go beyond the master equation approach addressing the question: how should we modify the way system and environment interact at a {\it microscopic} level in order to turn rare trajectories into typical as desired? On the other hand we go beyond continuous-time processes and address {\it discrete}-time quantum dynamics corresponding to a sequence of stochastic quantum maps on the open system.

To achieve the above, we use a quantum collision model (CM): in fact the natural and simplest microscopic framework for describing quantum trajectories and weak measurements \cite{brunSimple2002, altamirano_unitarity_2017, Gross2018, Ciccarello,Cilluffo2020Timebin, Seah_2019}. 
In a CM [see \fig1], the system of interest unitarily interacts, in a sequential way, with a large collection of environmental subunits (or probes), which constitute a thermal bath. After the collision, each probe undergoes a projective measurement, whose result is recorded. The sequence of measurement outcomes (see \fig1) defines a quantum trajectory. Since the unitary collision correlates the system and the probe, measuring the latter changes the state of the system as well. This change is tiny in most cases but occasionally can be dramatic and culminate in a quantum jump.

Exploiting thermodynamic functionals, we characterize the ensemble of trajectories in CMs and show how the system-probe interaction can be modified so as to bias the statistics of measurement outcomes on the probes. Notably, this unveils the physical mechanism turning rare trajectories into typical. As will be shown, for short collision times, the modified dynamics is obtained by adding extra collisions which enforce the system dynamics far from the average one so as to sustain a trajectory with desired measurement outcomes.

\section{Collision model}
\begin{figure}
	\includegraphics[scale=0.4,angle=0]{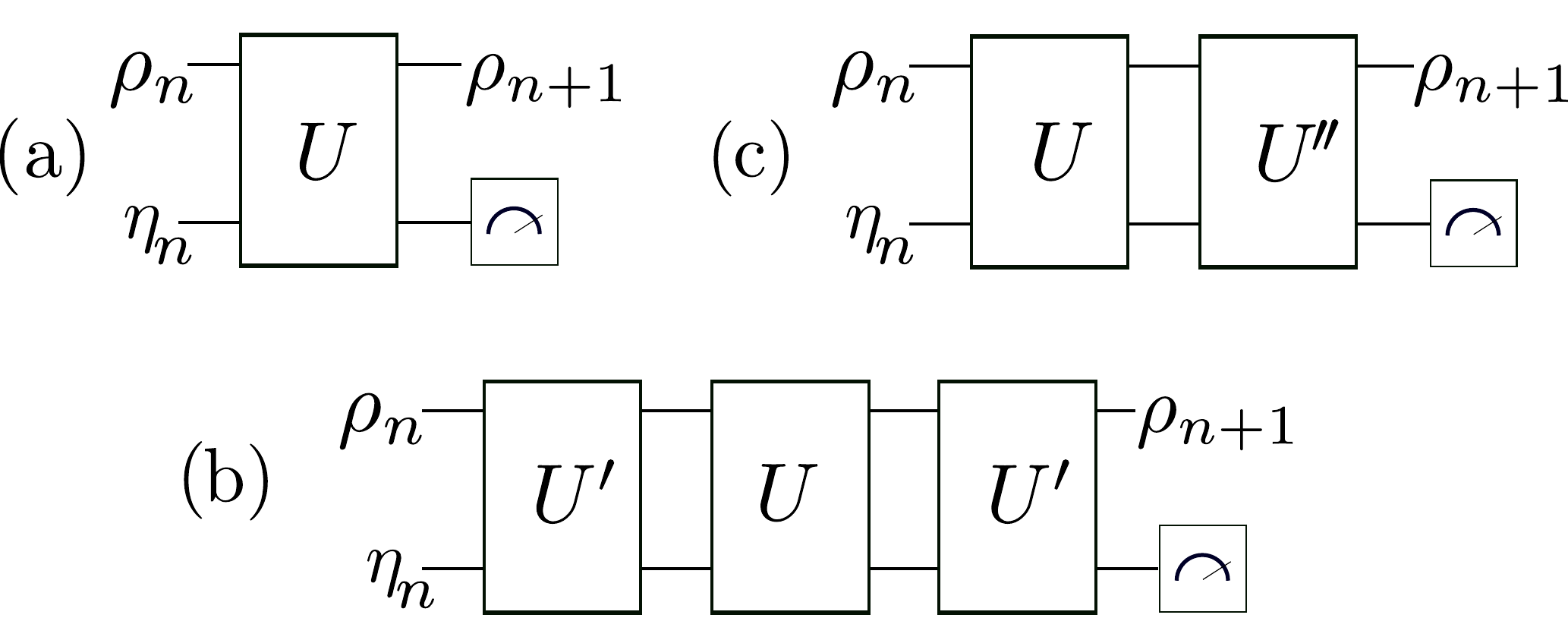}
	\caption{{\bf Quantum circuits.} (a): Quantum-circuit representation of a system-probe collision followed by a probe measurement. The system, whose state at the discrete time $n$ is given by $\rho_n$, collides with $n$th probe, initialized in the state $\eta_n$. The collision is unitary and implemented by the operator $U=U(H_S,V)$, which can be represented as a quantum gate. After the collision, measuring the probe returns the updated state of the system $\rho_{n+1}$ (by averaging over all measurement outcomes). (b): Modified collision turning rare trajectories into typical: a pair of extra gates $U'=U(H_S',V')$ are added to $U$ (before measurement). (c): Same modified collision as in (b) implemented through only one additional collision with unitary $U''=U(H_S'',V'')$.}
	\label{fig2}
\end{figure}

The environmental probes [see \fig\ref{fig1}] are labeled by $n=1,2,...,N$ and assumed to be non-interacting. Each is modeled as a qubit with states $\{\ket{0}, \ket{1}\}$ (note that quantum optics master equations and photo detection schemes are always describable in terms of qubit probes \cite{Wiseman, Gross2018}).
Each system-probe collision is described by the pairwise unitary
\begin{align}
U(H_S,V) =\exp[-i (H_S\otimes  \mathbb{1}{+}V) \Delta t)],
\label{U-coll}
\end{align}
with $H_S$ the free Hamiltonian of system $S$ (generally including a drive) and $V$ the $S$-probe interaction Hamiltonian. Note that $U$ can be seen as a gate acting on system and probe \cite{scarani_thermalizing_2002, Ciccarello} according to an associated quantum-circuit representation (see Fig.~\ref{fig2}).
In the following, we assume that initially $S$ and the probes are in the uncorrelated state $\varrho_0 = \rho_0 \bigotimes_n \eta_n$ with $\rho_0$ ($\eta_n$) the initial state of $S$ (probe $n$). We  will set $\eta_n=\ket{0}\!_n\langle 0|$ (the generalization to mixed state is straightforward). 

Right after colliding with $S$, under the action of the unitary $U(H_S,V)$, each probe is measured in the basis $\{\ket{k_n}\}$ with $k=0,1$ [see \fig\ref{fig2}(a)]. In an atom-field setup 
outcome $\ket{0}$ means no emission while $\ket{1}$ signals one photon emitted by $S$ and detected. The state of $S$ after $n$ steps, $\rho_n$, is the average over all possible discrete trajectories (unconditional dynamics). Between two  subsequent steps, it evolves as $\rho_{n+1}=\mathcal{E}[\rho_{n}]$, where the map
\begin{equation}
\mathcal{E}[\rho]:=\sum_{k=0}^1 K_k \rho K_k^\dag\, ,\quad \mbox{with} \quad  K_k=\langle k|U(H_S,V)|0\rangle\, ,
\label{K-map}
\end{equation}
is completely positive and trace preserving (CPT). $K_k$ are the so called Kraus operators acting on $S$. In particular, trace preservation (equivalent to probability conservation) holds due to  $\sum_{k=0,1} K^\dagger_k K_k=\mathbb{1}$.
 
We take the linear system-probe coupling 
\begin{align}
V=\tfrac{1}{\sqrt{\Delta t}} (J \otimes \sigma_{+}+J^\dag \otimes \sigma_{-})\,,
\label{int_lin}
\end{align}
where $J$ is an operator on $S$ having the units of the square root of a frequency, and 
$\sigma_{-}=\sigma_{+}^\dag=|0\rangle\!\langle 1|$. 
In spite of its simplicity, this model of interaction describes a wide variety of representative physical situations \cite{Gross2018, Doherty_1999}.
Also, note that \eqref{K-map} is independent of the probe label since so are $U$ and $\eta_n$.

Being a sequence of identical CPT maps, the overall discrete dynamics of $S$ is Markovian and, in the limit of vanishing collision time, it reduces to the continuous-time Markovian open quantum dynamics described by the Lindblad master equation $\dot \rho=\mathcal{L}[\rho]$ with \cite{Lindblad1976,Gorini1976}
\begin{equation}
\mathcal{L}[\rho]=-i\left[H_S,\rho\right]+J\rho J^\dagger -\tfrac{1}{2}\left\{J^\dagger J, \rho \right\}\, .
\label{Lindblad}
\end{equation}

\section{Biased collisional trajectories}
In contrast to the average (deterministic) dynamics generated by \eqref{K-map}, each specific quantum trajectory is composed of the specific measurement outcomes on the probes and is thus stochastic. At each step, the state of $S$ evolves as \cite{BrunPRA2000}
\begin{align}
\ket{\psi_{n+1}} = K_k \ket{\psi_n} / \|K_k \ket{\psi_n}\|
\end{align}
with $p_k =\| K_k \ket{\psi_n} \|^2$ being the probability to measure the $n$th probe in state $\ket{k}$ (we have assumed an initial pure state for the system, $\rho_0=|\psi_0\rangle\langle \psi_0|$, for the sake of argument). 

Each $K_k$ is in one-to-one correspondence with a particular measurement outcome.
 Here we focus on measurement outcomes described by operator $K_1$. Let then $P_N(M)$ be the probability of observing $M$ times the action of $K_1$ in a realization of the collision dynamics up to the discrete time $N$. For large $N$, this has  the form \cite{touchette2009large,garrahan2010thermodynamics,StegmannPRB2018,Chetrite_2015}
\begin{align}
P_N(M)\sim e^{-N\varphi(m)}\, ,
\end{align}
with $m=M/N$ being the frequency with which the probe has been measured in state $\ket{k=1}$. The (positive, semi-definite) function $\varphi$ is the so-called \textit{large deviation function}. It only vanishes when $m$ is equal to its typical value $\langle m\rangle$, i.e.,~the most likely to observe. This function fully characterizes the statistics of the random variable $M$. To obtain it, it is convenient to define the moment generating function (MGF) of the observable
\begin{align}
Z_N(s):=\sum_{M=0}^\infty P_N(M)e^{-s\, M}\xrightarrow[N \gg 1]{} e^{N \theta(s)} \, ,
\label{mgf}
\end{align}
where the real variable $s$ is called ``counting field", and 
$\theta(s)$ is the scaled cumulant generating function (SCGF) valid at stationarity for the observable $M$ \cite{touchette2009large} 
\begin{align}
\theta(s):=\lim_{N\to\infty}\frac{1}{N} \log Z_N(s),
\label{scgf}
\end{align}
In line with the arguments of \cite{touchette2009large,garrahan2010thermodynamics}, the SCGF can be calculated as the logarithm of the largest real eigenvalue of a \textit{tilted} Kraus map [\cf\eqref{K-map}] (see the Appendix \cite{SM} for further details).
\begin{align}
\mathcal{E}_s [X] = K_0 X K_0 + e^{-s} K_1 X K_1 \, .\label{Kraus-s}
\end{align}
The map $\mathcal{E}_s$ does not represent a physical process, but is rather a mathematical tool that is of help to recover $\theta(s)$. 
The (physical) process is retrieved for $s=0$.
The probability distribution $P_N(M)$ is determined by the behavior of $\theta(s)$ through derivatives with respect to $s$, taken at the ``physical point" $s=0$. 
Yet, looking at \eqref{mgf}, after normalizing by $Z_N(s)$, one can define a set of {\it biased} probabilities 
\begin{align}
P_N^s(M)=\frac{e^{-s\, M}P_N(M)}{Z_N(s)}\, .\label{PNMs}
\end{align}
For $s>0$, these probabilities enhance occurrence of trajectories featuring smaller-than-typical values of $M$, while for $s<0$, instead, larger values of $M$ are favored \cite{garrahan2011quantum}.
Remarkably, these apparently fictitious probabilities in fact describe rare ensembles of trajectories of the original collision model \cite{PhysRevLett.111.120601}. 
Cumulants of the biased probability distribution $P_N^s(M)$  can be determined through derivatives of $\theta(s)$ for values of $s$ different by zero; for instance, the rate of the measurement of probes in state $\ket{1}$ is, for $P_N^s(M)$, $\langle m \rangle_s=\tfrac{1}{N}\langle M \rangle_s = -\theta'(s)$.

\begin{figure}
	\includegraphics[scale=0.35,angle=0]{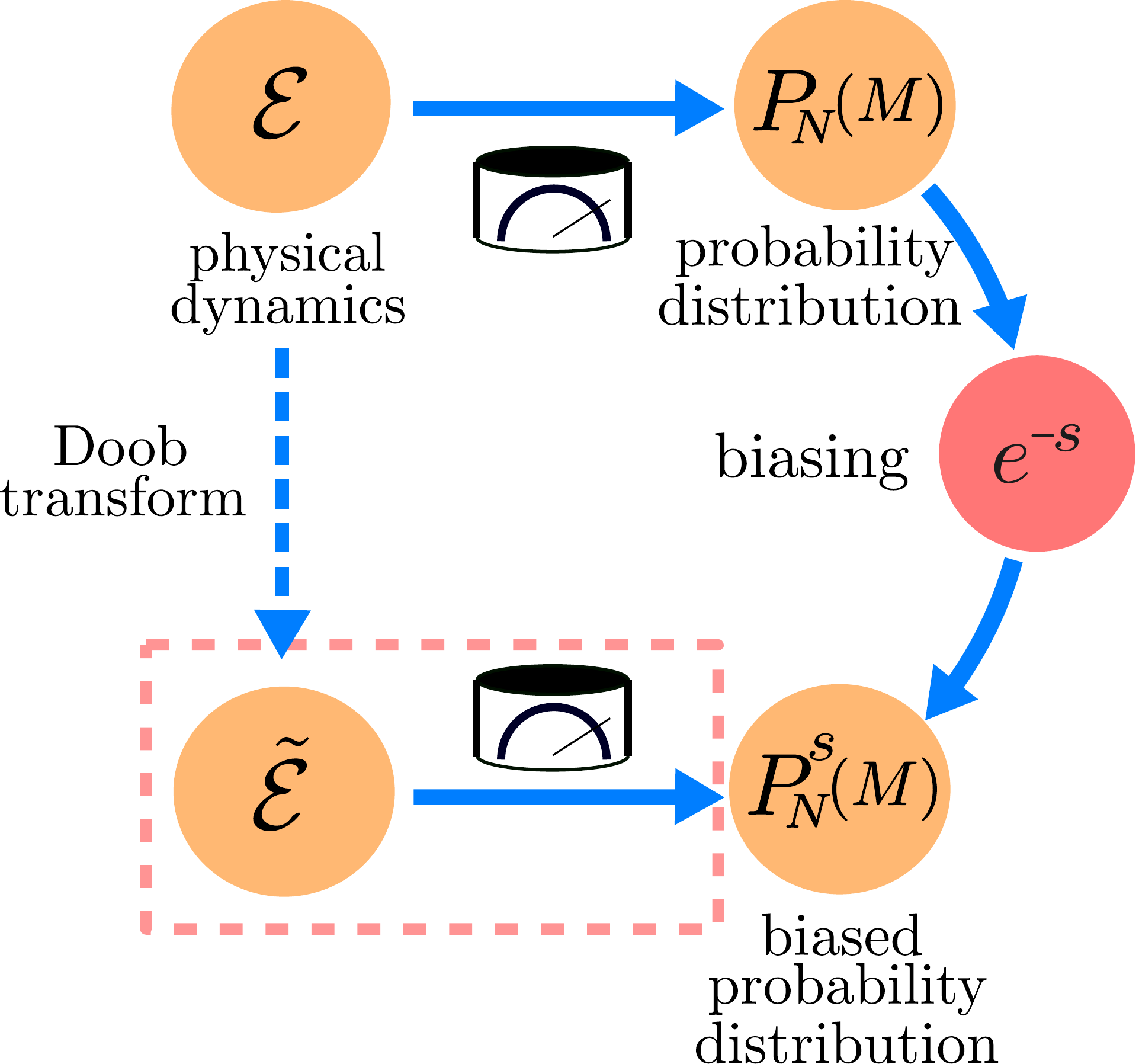}
	\caption{\textbf{Turning biased trajectories into typical.}
The collision model described by the dynamical map $\mathcal{E}$ is associated with the probability distribution $P_N(M)$ for probe measurements. The biased distribution $P_N^{s}(M)$ shows enhanced or suppressed typical values of $M$ as $s$ is varied. To control measurement outcomes of the collision model, one needs to engineer a process $\tilde{\mathcal{E}}$ (dashed box) associated with this probability distribution. We achieve this task by means of the Doob transform of the dynamics.
}
	\label{fig3bis}
\end{figure}

\section{Turning biased trajectories into typical}
So far we have constructed the probability distribution \eqref{PNMs} by hand and noted how these actually describe rare dynamical events. Here, we show how to modify the system-probe collision in a way that $P_N^s(M)$ become instead physical probabilities. In other words we will show how, by tuning the interaction between system and probes, the rare behavior of
the original process can become the typical one of the
new dynamics (see \fig\ref{fig3bis}).

\begin{figure*}
	\includegraphics[scale=0.5,angle=0]{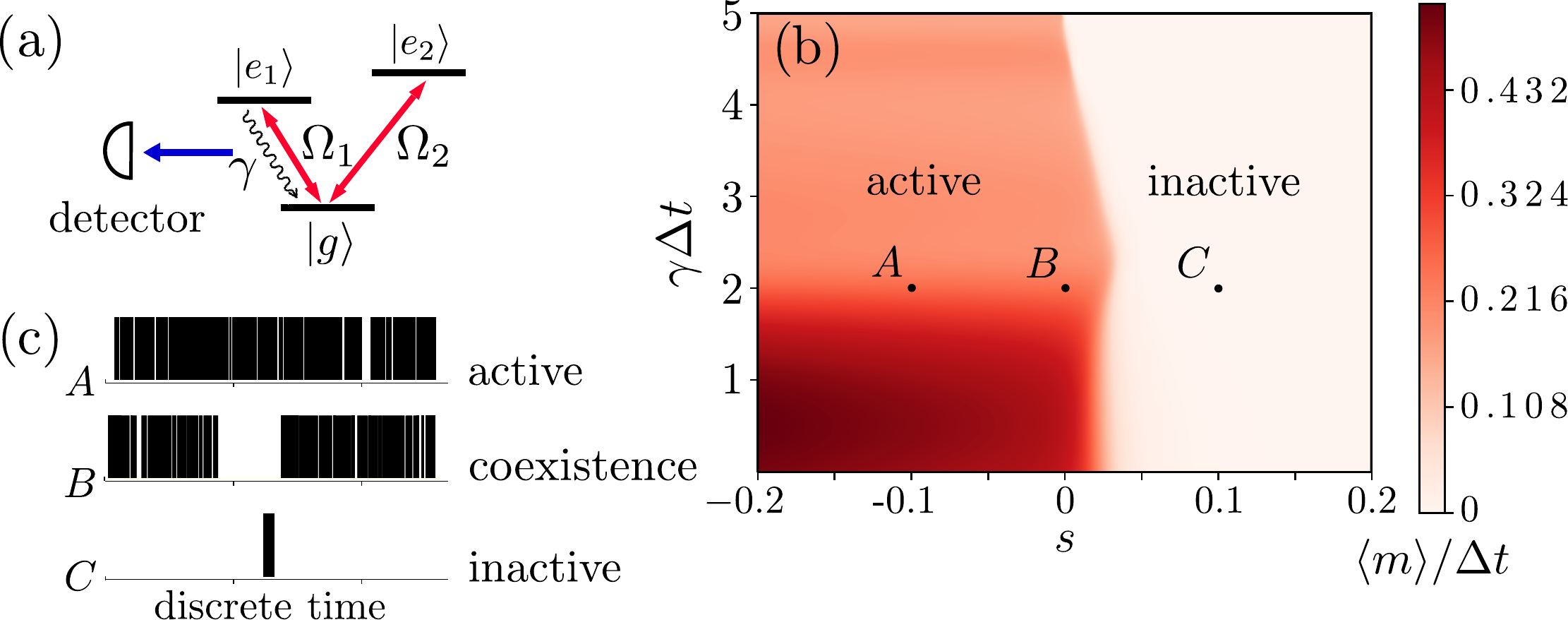}
	\caption{{\bf Discrete-time quantum trajectories of a three-level system.} (a) Level configuration: each transition $|g\rangle \leftrightarrow|e_k\rangle$ is driven with Rabi frequency $\Omega_k$ with $k=1,2$. During each unitary collision, a coherent exchange of excitations occurs between the probe and the system $S$. In particular, only transition $|g\rangle \leftrightarrow|e_1\rangle$ couples to the probe. 
Thus, measuring the probe in state $\ket{1_n}$ 
signals that the environment gained an excitation at the expense of the energy of the system
 $S$ which decays to the ground state $\ket{g}$ (emulating photon emission).
(b) Normalized activity (average emission rate divided by $\Delta t$) as a function of $s$ and $\Delta t$. The domain close to the boundary line separating active and inactive phases is a coexistence region. The variation of contrast for growing $\Delta t$ witnesses changes in sharpness of the active-inactive region crossover. (c) Sampled representative trajectories for a collision time $\gamma \Delta t \simeq 2$, with each tick recording a probe measurement in $\ket{1}$. Trajectories in the active phase (see A) show a dense emission of excitations from $S$ into the environment. In the inactive regime (C), instead, probes are rarely detected in $\ket{1}$.
Close to the boundary line between these two phases (see B), time intervals in which $S$ emits frequently are intermittent with intervals during which probes are almost never measured in $\ket{1}$.}
	\label{fig3}
\end{figure*}

As mentioned earlier $P_N^{{s}}(M)$ is generated by the tilted map $\mathcal{E}_{{s}}$ [\cf\eqref{Kraus-s}] which is {\it not} CPT (i.e., it does not represent a legitimate physical process) since probability is not preserved. The task is thus to turn $\mathcal{E}_{{s}}$ into a well-defined CPT map. This is achieved by introducing a \emph{Doob} transform of the dynamics \cite{10.1143/PTPS.184.304,garrahan2010thermodynamics, carollo2018making} for {\it discrete-time} quantum processes, embodied by the auxiliary CPT map (see the Appendix \cite{SM})
\begin{equation}
\tilde{\mathcal{E}}[X]=\tilde{K}_0X\tilde{K}_0^\dagger +\tilde{K}_1X\tilde{K}_1^\dagger\, .\label{doob-discrete}
\end{equation} 
The explicit expression of the modified Kraus operators $\tilde{K}$ is given in \cite{SM}.
The probability distribution associated with the map $\tilde{\mathcal{E}}$ is exactly the desired one $P_N^{{s}}$ for long times.

Notably, for short collision times $\Delta t$, the replacement $\mathcal{E}\rightarrow \tilde{\mathcal{E}}$ [\cf\eqref{K-map}] is equivalent to changing the system-probe collision unitary as
\begin{equation}
U({H}_S,{V})\rightarrow U(\tilde{H}_S,\tilde{V})\,.
\label{change}
\end{equation}
Here the new Hamiltonian $\tilde{H}_S$ and jump operator $\tilde{J}$ match those obtained via the Doob transform for continuous-time Lindblad processes \cite{garrahan2010thermodynamics,carollo2018making,PhysRevLett.122.130605}. As a consequence, the new Kraus operators are
\begin{align}
\tilde{K}_k=\langle k |U(\tilde{H}_S,\tilde{V})|0\rangle\,. 
\label{Kraus_D}
\end{align}
The new system-probe collision unitary \eqref{change} can be decomposed as (see the Appendix)
\begin{equation}
U(\tilde{H}_S,\tilde{V})= U(H'_S,V')U(H_S,V)U(H'_S,V')\,,\label{3coll}
\end{equation}
where $H'_S=\tfrac{1}{2}(\tilde{H}_S{-}H_S)$ and $V'=\tfrac{1}{2}(\tilde{V}{-}V)$. The associated quantum circuit is shown in \fig\ref{fig2}(b). This decomposition makes apparent the mechanism by which rare events can be sustained so as to make them typical: {\it extra} unitary collisions, added to the original one $U(H_S,V)$, drive the system away from typicality, pinning its dynamical behavior to the fluctuations of interest.

Note also that the same task can be accomplished by a single additional collision according to
\begin{equation}
U(\tilde{H}_S,\tilde{V})= U(H''_S,V'')\,U(H_S,V)\,
\label{2-coll1}
\end{equation}
with 
\begin{equation}
H''_S=2H'_S\,,\,\,\,V''=2V'+i\tfrac{\Delta t}{2}\,[\tilde{V},V]\,.
\label{2-coll}
\end{equation}
This is obtained from \eqref{3coll} by swapping the last two unitaries and applying the Baker-Campbell-Hausdorff formula \cite{greiner2013field} to leading order. 
Note that the second term in $V''$ (cf. \eqref{2-coll}) is of order $\mathcal{O}(1)$ in $\Delta t$, and represents an extra system-prob coupling. \eqs \eqref{2-coll1} and \eqref{2-coll} hold for any collision time $\Delta t$.

\section{Driven three-level system.} 
As an example, we discuss here a simple system, with rich dynamical behaviour which illustrates how our ideas can be exploited to investigate reduced-system discrete-time dynamics in metastable or prethermal regimes as well as to bias and drive such interesting dynamics.

Let $S$ be a coherently driven three-level system [see \fig\ref{fig3} (a)].
Each transition $|g\rangle \leftrightarrow|e_k\rangle$, with $k=1,2$, is driven with a Rabi frequency $\Omega_k$ according to the Hamiltonian
\begin{equation}
{H}_{S} = \sum_{k} \Omega_k (\s^{(k)}_+ + \s^{(k)}_- ) \,,
\end{equation}
where $\s^{(k)}_{-} = |g\rangle_S  \langle e_k| = \s^{(k)\,^\dag}_{+}$. For the sake of argument we assume the lasers to be in resonance with the atomic transitions.  Additionally, we set $J=\sqrt{\gamma}\,\s^{(1)}_-$ [\cf\eqref{int_lin}], meaning that only state $\ket{e_1}$ can decay with rate $\gamma$ by emitting an excitation into the environment (corresponding to outcome $\ket{1_n}$). 
For short collision times, intermittent emission is known to occur  \cite{PlenioJumpRMP,CookPRAIntermittency}, which can been explained as the coexistence of two deeply different phases of emission much like a first-order phase transition \cite{garaspects}.
Notably, the developed framework allows to investigate such transition-like behaviour away from the Lindblad dynamical regime, i.e., for {\it finite} collision times $\Delta t$.
The collision Hamiltonian reads
\begin{align}
V=\sqrt{\tfrac{\gamma}{\Delta t}} (\s^{(1)}_- \otimes \sigma_{+} + \s^{(1)}_+ \otimes \sigma_{-})\,,
\label{int_ex}
\end{align}
thus through the biased map $\mathcal{E}_s$ (\cf\eq \eqref{Kraus-s}) we work out the auxiliary map $\tilde{\mathcal{E}}$ and study the statistics of the quantum trajectories generated by quantum jump MonteCarlo.
To this end, we plot in \fig\ref{fig3}(b) the time-averaged rate of probe measurements in state $\ket{1}$, $\langle m\rangle/\Delta t = -\partial_s (\theta(s,\Delta t))/\Delta t$, as a function of $s$ and $\Delta t$ for $\Omega_1/\gamma=1$ and $\Omega_1/\Omega_2=1/10$.
This dynamical order parameter allows us to distinguish active (bright) and inactive (dark) trajectory regimes [some representative samples of quantum trajectories are shown in \fig \ref{fig3}(c)]. 
The boundary line --clearly visible in \fig\ref{fig3}(b)-- represents a sharp crossover between the two dynamical regimes. Along this boundary, trajectories feature intermittent emission of excitations from the system. 
As $\Delta t$ grows up, the crossover occurs at a different value of $s$ and its sharpness changes. Thus, away from the short-$\Delta t$ (Lindblad) regime, both typical and atypical emission rates are modified.

\section{Conclusions} 
We presented a microscopic framework for the statistical characterization of quantum trajectories in discrete-time processes. 
This provides a quantitative tool for studying dynamical fluctuations beyond the standard continuous-time regime corresponding to the Lindblad master equation.
A recipe was given allowing to turn rare quantum trajectories into typical upon addition of extra collisions between the system and each probe.
It is worth noting that this is reminiscent of a giant-atom dynamics (a giant atom couples to the field at two or more points \cite{Kockum5years}), which can indeed be described as  cascaded collisions \cite{giovannettiMaster2012a,giovannettiMaster2012,LorenzoPRAflux} yet involving the same system $S$ \cite{Carollo2020CCC}. \\
While we have focussed on collisions of the form described in \eqref{U-coll} and \eqref{int_lin}, our results for discrete-time collision models do not depend on the specific form of the collision unitary. We note that also the interpretation of the Doob dynamics as a collision model with an additional collision should extend straightforwadly to such more general cases \cite{SM}.

We note that it is also possible to use this formalism to obtain the finite-time statistics of emissions for discrete-time quantum maps as well as their finite-time Doob transform. This can be done by following essentially the same steps used for the continuous-time case, as for instance done in \cite{carollo2018making} where, in order to obtain the finite-Doob dynamics, the  continuous-time dynamics has been first discretized.

The method introduced here shows how to engineer open quantum dynamics in order to produce desired emission patterns, without the need for changing the detection/ post-selection scheme \cite{BudiniPRE2011}. 
Moreover, the presented qubit-based protocol can be implemented with experimental quantum simulator platforms based on trapped ions \cite{Schindler_2013} or Rydberg atoms \cite{Browaeys_2020,Weimer_2010}.

\section{Acknowledgments}
F. Carollo~acknowledges  support  through  a  Teach@T\"ubingen Fellowship. IL~acknowledges  support from EPSRC  [Grant  No.~EP/R04421X/1], from The Leverhulme Trust  [Grant No.~RPG-2018-181] and from the ``Wissenschaftler-R\"uckkehrprogramm GSO/CZS" of the Carl-Zeiss-Stiftung and the German Scholars Organization e.V.. 
A. Carollo acknowledges support from the Government of the Russian Federation through Agreement No.~074-02-2018-330 (2).
We acknowledge support from MIUR through project PRIN Project 2017SRN-BRK QUSHIP.
The research leading to these results has received funding from the European Union’s H2020 research and innovation programme [Grant Agreement No. 800942 (ErBeStA)].
\bibliographystyle{apsrev4-1}
\bibliography{biblio}

\onecolumngrid
\newpage

\renewcommand\thesection{A\arabic{section}}
\renewcommand\theequation{A\arabic{equation}}
\renewcommand\thefigure{S\arabic{figure}}
\setcounter{equation}{0}
\setcounter{figure}{0}

\appendix

\section*{Appendix}

\subsection{Doob transform of the discrete process.}

Although the large deviation function $\varphi(m)$ encompasses full information about the asymptotic behavior of the probability distribution $P_N(M)$, it is not, in general, easy to access by direct calculation. 
The diagonalization of tilted map $\mathcal{E}_{s}$ (\cf \eqref{Kraus-s}) does the job, providing the SCGF $\theta(s)$ (\cf \eqref{scgf}) as the logarithm of its maximum eigenvalue that we name $\Lambda_s := e^{\theta(s)}$. 
$\theta(s)$ captures the asymptotic behavior of cumulant generating function $\log Z_N(s)$, and is linked to $\varphi(m)$ through the Legendre-Fenchel transform \cite{touchette2009large}
\begin{align}
\varphi(m)=-\max_{\forall s}\left[m\, s + \theta(s)\right]\, .
\end{align}

Furthermore $\mathcal{E}_{s}$ biases the original probabilities $P_N(M)$ through an exponential factor.
By exploiting the partition function (MGF) $Z_N(s)$ we can thus define the tilted probability
\begin{align}
P_N^{s}(M)=\frac{e^{-s M}P_N(M)}{Z_N(s)}\, .
\end{align}
This ensemble --so-called $s$-ensemble-- contains information about the properties and the dynamical features associated with a rare event of the originial process. However, $\mathcal{E}_{s}$ is not a well-defined quantum discrete dynamics since the dual map does not preserve the identity, $\mathcal{E}_{s}^*[{\mathbb 1}]\neq{\mathbb 1}$. 

Nonetheless, as we now demonstrate, it is possible to transform this tilted into a proper dynamics, which reproduces as typical the rare outcomes of the original collision model $\mathcal{E}$. We can obtain this dynamics as follows. Suppose we are interested in the behaviour of the system associated with probabilities $P_N^{s}(M)$. Then, we can define the (Doob) quantum discrete quantum 
\begin{equation}
\tilde{\mathcal{E}}[\rho]=\sum_{k=0,1}\tilde{K}_k \rho \tilde{K}_k^\dagger \, , \quad \mbox{ where }\quad  \tilde{K}_0 =\frac{1}{\Lambda_s^{1/2}}\ell^{1/2}K_0\ell^{-1/2} \qquad \mbox{ and } \quad \tilde{K}_1 =\frac{e^{-s/2}}{\Lambda_s^{1/2}}\ell^{1/2}K_1\ell^{-1/2}\, .
\label{discrete-Doob}
\end{equation}
Here we have that $\ell$ is the left eigen-operator of the tilted map $\mathcal{E}_s$ associated with its eigenvalue with largest real part $\Lambda_s = e^{\theta(s)}$. Namely, $\ell$ is the operator such that 
\begin{align}
\mathcal{E}_s^*[\ell]=\Lambda_s \, \ell\, .
\end{align}
The map $\tilde{\mathcal{E}}$ is completely positive and we also have that $\tilde{\mathcal{E}}^*[{\mathbb 1}]={\mathbb 1}$. The latter equality follows from
\begin{align}
\tilde{\mathcal{E}}^*[{\mathbb 1}]=\frac{1}{\Lambda_{s}}\ell^{-1/2}\, \mathcal{E}_s^* [\ell]\ell^{-1/2}=\frac{1}{\Lambda_s}\ell^{-1/2}\left(\Lambda_{s}\ell\right)\ell^{-1/2}={\mathbb 1}\, .
\end{align}
Because of these properties, the map in \eqref{discrete-Doob} is a proper discrete quantum dynamics and, as we have discussed, reproduces as typical the rare event of the original processes  $P_N^{s}(M)$. 

\subsection{Doob transform in the collision model.}
In the above section, we have demonstrate how to obtain the Doob dynamics of a discrete-time quantum process. Here, instead, we want to consider that our initial dynamics describes a collision model, meaning that $\Delta t\to 0$. Using this fact, we show how the Doob dynamics is in fact a new collision model with effective Hamiltonian and jump operator which coincides with those of the Doob transform for the continuous-time Lindblad case \cite{carollo2018making}. 
First of all, we notice that in the collision model limit, $\Delta t\ll1$, the tilted Kraus map is approximately given by 
\begin{align}
\mathcal{E}_s[\rho]\approx e^{\Delta t \, \mathcal{L}_s} [\rho]\, ,
\end{align}
where $\mathcal{L}_s$ is the tilted Lindblad operator \cite{garrahan2010thermodynamics,carollo2018making}
\begin{align}
\mathcal{L}_{s}[\rho]=-i[H_S,\rho]+e^{-s}J\rho J^\dagger -\frac{1}{2}\left\{\rho,J^\dagger J\right\}\, . 
\end{align}

As such, the left eigen-operator of $\mathcal{L}_s$ is approximately also the eigen-operator of $\mathcal{E}_{ s}$, $\ell$, at first-order in $\Delta t$. This also implies that the largest real eigenvalue of the tilted map can be written as
\begin{align}
\Lambda_s\approx e^{\Delta t\, \chi(s)}\, ,
\end{align}
where $\chi(s)$ is given by the largest real eigenvalue of the tilted Lindbladian map $\mathcal{L}_s$. 

We can now focus on the Doob transform in \eqref{discrete-Doob} and consider the small collision-time limit. The second term on the right hand side is thus equivalent to
\begin{align}
\tilde{K}_1\rho \tilde{K}_1^\dagger \approx \frac{e^{-s}}{e^{\Delta t\, \chi(s)}}\Delta t\, \ell^{1/2} J \ell^{-1/2}\, \rho \, \ell^{-1/2} \tilde J^\dagger \ell^{1/2} \approx \Delta t\, \tilde J \rho \tilde J^\dagger\, ;
\end{align}
here $\tilde{J}=e^{-s/2}\ell^{1/2}J\ell^{-1/2}$ (which corresponds to the jump operator of the continuous time Doob dynamics \cite{carollo2018making}) and the last term ${e^{\Delta t\, \chi(s)}}$ only contributes at the zero-th order in $\Delta t$. 

Considering the first term on the right hand side of \eqref{discrete-Doob}, up to first order in $\Delta t$, we obtain 
\begin{align}
\tilde{K}_0\rho\tilde{K}_0^\dagger \approx 1+\left[-i\ell^{1/2}H_{\rm eff}\ell^{-1/2}\rho+i\rho\ell^{-1/2}H_{\rm eff}^\dagger \ell^{1/2}-\chi(s)\rho\right]\Delta t\, ,
\end{align}
and this, with similar computation as those done in Ref.~\cite{carollo2018making} gives 
\begin{align}
\tilde{K}_0\rho\tilde{K}_0^\dagger \approx 1 -i\left(\tilde{H}_S-\frac{i}{2}\tilde{J}^\dagger \tilde{J} \right)\rho\Delta t+i \rho \left(\tilde{H}_S+\frac{i}{2}\tilde{J}^\dagger \tilde{J}\right)\Delta t\, ,
\end{align}
where $\tilde{H}_S$ coincide with the Hamiltonian of the continuous time Doob dynamics
\begin{align}
\tilde{H}_S &= \frac{1}{2} \ell^{1/2} \left(
H  - \frac{i}{2} J^\dag J 
\right) \ell^{-1/2} + {\rm H.c.}\,.
\end{align}

In light of this result, we can write the unitary interaction between system and probe as a new collision model with
\begin{align}
U(\tilde{H}_S,\tilde{V})=\exp[-i (\tilde{H}_S\otimes {\mathbb 1}{+}\tilde{V}) \Delta t)]\, ,
\end{align}
and 
\begin{align}
\tilde{V}=\frac{1}{\sqrt{\Delta t}} (\tilde{J}\otimes \sigma_{+}+\tilde{J}^\dag \otimes \sigma_{-})\, .
\end{align}

\subsection{The Doob collision dynamics as a three-collision model.}
In this section we show how it is possible to write the Doob dynamics as a collision model where system and probe collide three times with one collision being exactly the original one.

To show this we start by writing the Doob unitary collision with the $n$-th ancilla as
\begin{align}
U(\tilde{H}_S,\tilde{V})=\exp\left[-i(A+B)\right]\, ,
\end{align}
where we have 
\begin{align}
A=\Delta t\left(H_S\otimes {\mathbb 1}+V\right)\, , \qquad \mbox{and}\qquad B=\Delta t\left(\tilde{H}_S\otimes{\mathbb 1}-H_S\otimes{\mathbb 1}+\tilde{V}-V\right)\, .
\end{align}
Due to the fact that we are interested in the regime $\Delta t\ll1$, we just need to preserve the terms of the unitary operator $U(\tilde{H}_S,\tilde{V})$ only up to first order in $\Delta t$. Recalling that terms $V \Delta t$ are actually of order $\sqrt{\Delta t}$ this means it is sufficient to guarantee that the unitary $U(\tilde{H}_S,\tilde{V})$ is preserved up to the second order products of $A,B$. A possible decomposition is thus given by the second-order Trotter decomposition
\begin{align}
U(\tilde{H}_S,\tilde{V})=\exp(-iB/2)\exp(-iA)\exp(-iB/2)+o(\Delta t)\, .
\end{align}
Noticing that $\exp(-iA)=U(H_S,V)$, and since we can define  
\begin{align}
\exp(-iB/2)=U(H'_S,V')\, , \quad \mbox{ with } \quad H'_S=(\tilde{H}_S-H_S)/2\, \quad \mbox{ and }\quad V'=(\tilde{V}-V)/2\, ,
\end{align}
we are allowed to write the Doob collision as
\begin{align}
U(\tilde{H}_S,\tilde{V})\approx U(H'_S,V')U(H_S,V)U(H'_S,V')\, .
\end{align}

\subsection{Doob dynamics as a two-collision  model for finite collision time}
In this section we show that, also for the case of finite collision time, it is possible to have an interpretation of the Doob dynamics in \eqref{doob-discrete}, as a collision dynamics with one additional unitary collision. 

As shown in the main text, the Doob dynamics for generic discrete-time processes is  given by 
$$
\tilde{\mathcal{E}}[X]=\sum_{k=0,1}\tilde{K}_kX\tilde{K}_k\, .
$$
By Stinespring dilation theorem, it is possible to interpret these operators as 
$$
\tilde{K}_k=\bra{k}\tilde{U}\ket{0}\, ,
$$
where $\tilde{U}$ is a suitable unitary collision between system and bath. With $U$ being the original collision of the process, one can always write
$$
\tilde{U}=W_1U=UW_2\, ,
$$
where $W_1=\tilde{U}U^\dagger$ and  $W_2=U^\dagger\tilde{U}$. This means that we can interpret the unitary interaction of the Doob process $\tilde{U}$, as a the sequence of two collisions involving the original one and an extra one, which is sustaining as typical the rare behaviour of the original dynamics.

\end{document}